\documentclass[runningheads]{llncs}
\usepackage{graphicx}
\usepackage{amsmath,amssymb}
\usepackage[dvipsnames]{xcolor}
\usepackage{multirow}

\newcommand{\bm}[1]{\boldmath{#1}}

\newcommand{\HS}{\text{H}}
\newcommand{\softplus}[1]{\text{plus}_{\epsilon}(#1)}
\newcommand{\fHS}{\text{H}_{\sigma}}

\begin{document}
\title{Generative Modelling of 3D in-silico Spongiosa with Controllable Micro-Structural Parameters}
\titlerunning{Modelling of 3D Spongiosa with Controllable Micro-Structural Parameters}
\author{Emmanuel Iarussi\inst{1,2} \and Felix Thomsen\inst{1,3} \and Claudio Delrieux\inst{1,3}}
\authorrunning{Iarussi, Thomsen et al.}
\institute{Consejo Nacional de Investigaciones Cient\'ificas y T\'ecnicas, Argentina \and
Universidad Tecnol\'ogica Nacional, Facultad Regional Buenos Aires, Argentina \and
Universidad Nacional del Sur, Bah\'ia Blanca, Argentina\\
Corresponding address: \email{felix.thomsen@uns.edu.ar}}
\maketitle

\begin{abstract}
Research in vertebral bone micro-structure generally requires costly procedures to obtain physical scans of real bone with a specific pathology under study, since no methods are available yet to generate realistic bone structures \emph{in-silico}. 
Here we propose to apply recent advances in generative adversarial networks (GANs) to develop such a method. 
We adapted style-transfer techniques, which have been largely used in other contexts, in order to transfer style between image pairs while preserving its informational content.
In a first step, we trained a volumetric generative model in a progressive manner using a Wasserstein objective and gradient penalty (PWGAN-GP) to create patches of realistic bone structure \emph{in-silico}. 
The training set contained $7660$ purely spongeous bone samples from twelve human vertebrae (T12 or L1) with isotropic resolution of $164~\mu$m and scanned with a high resolution peripheral quantitative CT (Scanco XCT). 
After training, we generated new samples with tailored micro-structure properties by optimizing a vector $\boldsymbol{z}$ in the learned latent space. 
To solve this optimization problem, we formulated a differentiable goal function that leads to valid samples while compromising the appearance (content) with target 3D properties (style). 
Properties of the learned latent space effectively matched the data distribution.
Furthermore, we were able to simulate the resulting bone structure after deterioration or treatment effects of osteoporosis therapies based only on expected changes of micro-structural parameters.   
Our method allows to generate a virtually infinite number of patches of realistic bone micro-structure, and thereby likely serves for the development of bone-biomarkers and to simulate bone therapies in advance.
\keywords{Bone micro-structure \and progressive generative adversarial network \and structural morphing \and style-transfer \and XCT}
\end{abstract}

\section{Introduction}
The development of new methods for characterizing the micro-structure of spongeous bone is an active research field in constant progress.
For the development and analysis of specific structural parameters (e.g. bone volume ratio, trabecular separation or plate-to-rod ratio), often very simple \emph{in-silico} bone models are used~\cite{Moreno2012evaluation,Thomsen2016LFD} containing only few rods and plates intersecting each other, since they allow easiest control of the desired output.
These very simple models reflect rather poorly the real structure of bone.
A first attempt to generate more realistic bone has recently been presented, evolving a 3D structure from separated 2D slices, generated with a technique in Fourier domain~\cite{pena2019development}. 
The authors reported high accordance between generated and real samples regarding trabecular thickness, however further parameters have not been evaluated.
In this work we developed a direct method to create 3D micro-structural bone samples \emph{in-silico} that 1) contain realistic structures, and 2) allow to steer the properties to simulate changes of micro-structural parameters from deterioration or medical bone treatment.

In order to achieve these goals, we first trained a generative volumetric convolutional neural network (CNN) on isotropic patches of $32\times32\times32$ voxels (box of $5$mm diameter) to generate \emph{in-silico} HR-pQCT patches. 
Working in 3D severely increases the complexity of the generative models with respect to images (2D) or videos (2D+t).
Therefore we adapted a WGAN-GP architecture~\cite{gulrajani2017improved} to work with 3D data and trained it by progressively growing the resolution up to the target shape~\cite{karras2017progressive}, accordingly we call  this architecture {\em progressive} WGAN-GP (PWGAN-GP).
The latent vector $\boldsymbol{z}$ corresponds to a random point in a 32-dimensional hypersphere, which defines entirely the generated volume that the discriminator network is intended to differentiate from a real sample. 
\begin{figure*}[t!]
	\centering
	\includegraphics[width=\linewidth]{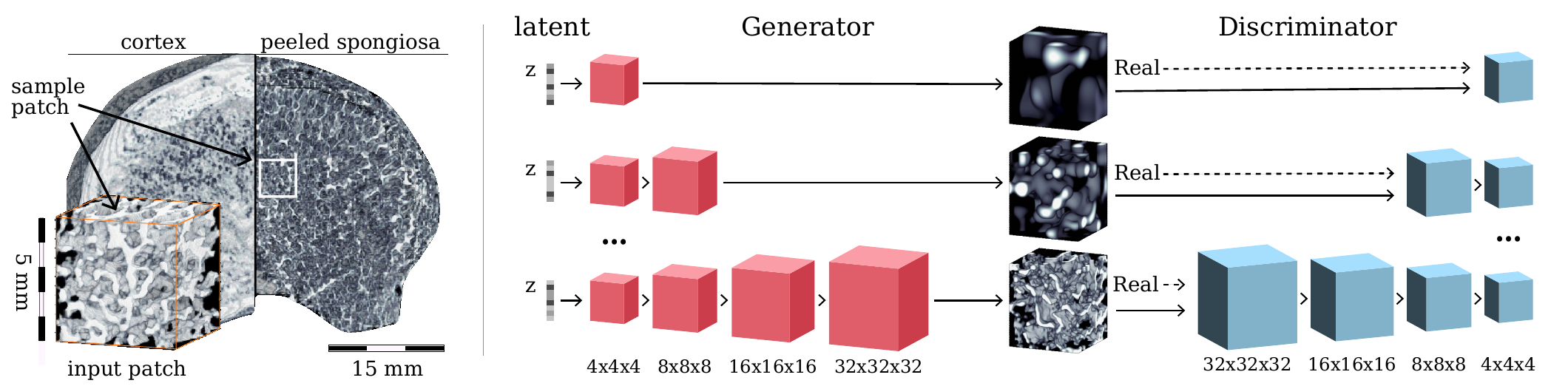}
	\caption{Left: 7660 purely spongeous HR-pQCT patches of $32^3$ voxels were sampled from twelve human vertebrae phantoms. Right: Two 3D Convolutional Neural Networks, Generator and Discriminator, were progressively trained to mimic the vertebrae volume distribution at incremental resolutions.}
	\label{fig:overview}
\end{figure*}

A key factor of our approach is to provide control over the generated samples, to be able to navigate in the learned latent space and to customize the morphological properties of the output volumes. 
Instead of shaping the latent space by means of a conditional GAN~\cite{mirza2014conditional}, conditioned on target micro-structural parameters, we took advantage of generative style transfer techniques~\cite{gatys2015neural} largely used in photo manipulation applications and other types of art-work.
In contrast to conditional GANs, this two-step approach, GAN and style transfer, does not require to retrain the generative model if the user desires to incorporate new control properties, i.e. a new set of micro-structural parameters. 
The style transfer mechanism allows to steer the properties and to simulate changes of micro-structural parameters, for instance from deterioration or medical bone treatment.
This is achieved by mapping the latent vector of an original sample to a vector of new micro-structural parameters (style) but still with a similar micro-structure (content), hence by solving an optimization problem with two conditions.
Thus, we show that not only neural networks but also style transfer strategies can be employed to generate and control structural properties in 3D bone CT scans. 
To the best of our knowledge, our work is the first attempt to address micro-structural customization of 3D volumes formulating it as a style transfer problem and also the first generative adversarial method to synthesize bone micro-structure~\cite{sorin2020creating}. 
Code, data and trained models will be released at \url{github.com/emmanueliarussi/generative3DSpongiosa}.

\section{Methods}
In this chapter we describe the examined generative models and the neural style transfer function. 
We adapted a 2D state of the art generative network model~\cite{karras2017progressive} and trained it with HR-pQCT volumetric scans of human vertebrae (Fig.~\ref{fig:overview}).
Once trained, we assessed the suitability of generated volumes qualitatively via 3D renderings, and quantitatively with standard metrics of trabecular bone.
Next, we generated samples with tailored micro-structural properties using a style transfer perspective over the learned latent space.
Synthetic samples contained new micro-structural properties but preserved the general content of the original sample. 

\subsection{Examined Deep Generative Models}\label{Sec:Deep_Volumetric_Generative_Model}
Generative adversarial networks (GANs)~\cite{goodfellow2016deep} are trained to stochastically generate samples close to a distribution represented by the training set. 
Despite the high success of these network architectures, training is still a very unstable process.
Therefore, several alternatives have emerged to deal with training issues.

In particular, we based our framework on Wasserstein GANs~\cite{arjovsky2017wasserstein}, consisting of a generator network $G~:~\boldsymbol{z}~\mapsto~\tilde{\boldsymbol{x}}$, and a discriminator network (also called \emph{critic} in the context of WGANs) $D~:~x~\mapsto~D(x)~\in~\mathbb{R}$ which were simultaneously trained to try to fool each other: while $G$ learned to generate fake samples $\tilde{\boldsymbol{x}}$ from an unknown distribution or noise $\boldsymbol{z}$, the critic $D$ learned meanwhile to distinguish fake from real samples. 
Formally, the training objective function optimizes: 
\begin{equation}
    \left. \min _{G} \max _{D \in \mathcal{D}} \underset{\boldsymbol{x} \sim \mathbb{P}_{r}}{\mathbb{E}}[D(\boldsymbol{x})]-\underset{\tilde{\boldsymbol{x}} \sim \mathbb{P}_{g}}{\mathbb{E}}[D(\tilde{\boldsymbol{x}}) ], \right.
    \label{eq:wgan_obj}
\end{equation}
where $\mathcal{D}$ is the set of 1-Lipschitz functions, $\mathbb{P}_{r}$ the data distribution and $\mathbb{P}_{g}$ the model distribution defined by $\tilde{\boldsymbol{x}}=G(\boldsymbol{z}), \boldsymbol{z} \sim p(\boldsymbol{z})$. 
Since enforcing the Lipschitz constraint is not trivial, we applied a gradient penalty mechanism (PG)~\cite{gulrajani2017improved} at every iteration with a critic parameter update (WGAN-PG).

We employed three additional non-progressive GANs to compare performance.
1) We trained the exact same architecture as described before but with an adversarial loss~\cite{goodfellow2016deep} and without extra regularization terms, gradient penalty and critic drift.
Instead, we added a sigmoid layer at the final discriminator step in order to output labels in range $[0,1]$. 
2) As a more advanced alternative we applied the full framework with gradient penalty and critic drift but without progressive training (WGAN-GP).
3) We additionally tested an alternative WGAN enforcing Lipschitz constraints by means of weight clipping. 
However, we omitted the third network from the detailed analysis since it suffered from mode collapse: all generated samples resulted extremely similar and did not resemble any realistic bone structure. 
All hyperparameters (total epochs, training rate, etc.) were kept fixed among all methods.

\subsection{Generating Samples with Custom Properties}
We framed the problem of generating samples with tailored micro-structural properties by formulating an optimization problem over the latent space of our generative model.
In the spirit of style-transfer techniques, we defined the \emph{content} $\boldsymbol{x_{t}}$ as the target volume we want to stick to.
The target volume could be a sample from our training data set, or also be produced by the generator network,
in that case $\boldsymbol{x_{t}}= G(\boldsymbol{z}_{t})$. 
A key contribution of our approach is to redefine the notion of \emph{style} for 3D scans of human vertebrae. 
If $\boldsymbol{x_{t}}$ provides the overall volume constraint, the style is given by a set of micro-structural properties $\boldsymbol{w_{t}}$, our generated sample has to satisfy.
Formally, we set up an optimization problem over $\boldsymbol{z}^{\prime}$ minimizing the sum of two conditions:
\begin{equation}
\min_{\boldsymbol{z}^{\prime}} \mu \|\boldsymbol{x_{t}}-G\left(\boldsymbol{z}^{\prime}\right)\|_{2}^{2} + \|\boldsymbol{w_{t}}- P(G(\boldsymbol{z}^{\prime}))\|_{2}^{2},
\label{eq:opt_style}
\end{equation}
with $\mu\geq0$ and $P(\cdot)$ the algorithm to compute a vector of differentiable micro-structural properties as given by $\boldsymbol{w_{t}}$, Sec.~\ref{Sec:Evaluation_of_Neural_Networks}.
The objective function is differentiable and can be minimized using gradient descent. 
Notice that the term accounting for the content is hard to minimize and that the global minimum may be not unique.
In practice, we set $\mu$ to a small value ($e^{-4}$) and optimized Eq.~\ref{eq:opt_style} with the Limited Memory Broyden–Fletcher–Goldfarb–Shanno algorithm (L-BFGS). 
As we show in the Results section, this optimization allows us to navigate the latent space learned by our network, retrieving plausible synthetic samples with the desired micro-structural properties. 

\section{Experimental Setup}
In this section we describe the sampling procedure, choice of hyperparameters and the applied statistics.

\subsection{Training Dataset}
Twelve human vertebrae (T12 and L1) were embedded into epoxy resin without damaging any trabeculae to become cylindrical vertebrae phantoms.
These phantoms were scanned on a high-resolution peripheral QCT (HR-pQCT) with isotropic resolution of $82~\mu$m, $59.4$~kVp and $900~\mu$As (Xtre\-me\-CT I, Scanco Medical AG, Br\"uttisellen, Switzerland) and automatically calibrated to density values.
The spongiosa has been peeled from the cortex with a semi-automatic procedure~\cite{Thomsen2016SI} and down-sampled to an isotropic resolution of $164~\mu$m, thereby increasing the signal-to-noise ratio and obtaining a lower voxel number per patch but still keeping most structural information (see Fig.~\ref{fig:overview}).

We defined then 7660 purely spongeous patches on the entire set of vertebra phantoms with isotropic size of $32\times32\times32$ voxels (box of diameter $5$~mm) and regular offset $8$ voxels in all directions ($1.3$ mm). 
Patches were normalized from $[-350,1100]~\text{mg/cm}^3$ to $[-1,1]$ (only $0.001\%$ or 1764 from 152 million voxels were thereby clamped), and denormalized again in production mode to compute structural parameters correctly.
Data have been augmented to 122,560 patches by employing all possible 16 axis-aligned rotations and reflections without imposing a misalignment of the vertical axis. 
This procedure avoids any interpolation artifact and respects the preferential structural and load orientation of bone.

\subsection{Parameter Settings}\label{sec:parameters_settings}
We progressively trained our network in four stages with isotropic samples of $4^3$, $8^3$, $16^3$ and $32^3$ voxels.
Each training stage consisted on five training epochs followed by five blending epochs to fade smoothly in the new layers. 
We used a batch size of $16$ in all stages and the ADAM optimizer~\cite{kingma2014adam} with $\beta_{1}=0$, $\beta_{2}=0.99$ and learning rate $=0.001$ for the generator and critic networks. 
We set the number of critic iterations per generator iteration to 3 and used gradient penalty with $\lambda=10$.
In order to further regularize the discriminator, we penalized outputs drifting away from 0 by adding the average of the critic output squared to its loss with weight $\epsilon_{drift}=0.001$.
Instead of batch normalization we used pixel normalization~\cite{karras2017progressive} after each 3D convolutional layer in the generator network.
Additionally, we updated the generator weights using Exponential Moving Average (EMA)~\cite{yazici2018unusual}.
In total, our generator and discriminator networks contained more than 200,000 trainable parameters each.

\subsection{Quantitative and Qualitative Evaluation}\label{Sec:Evaluation_of_Neural_Networks}
We implemented a set of commonly applied bone micro-structural parameters in Python: Bone mineral density (BMD), standard deviation of density values (BMD.SD), bone volume ratio (BV/TV), tissue mineral density (TMD), mean intercept length (MIL), parallel plate model dependent trabecular separation (Tb.Sp) and thickness (Tb.Th)~\cite{Thomsen2017PHD}.  
The vector $P(G(\boldsymbol{z}^{\prime}))$ as defined in Eq.~\ref{eq:opt_style} contained BMD, BV/TV, TMD and BMD.SD, that were specifically implemented as differentiable PyTorch functions, similar techniques have been used elsewhere~\cite{karam2019fast}.
Since micro-structural parameters were not independent from each other, either for physical (e.g. BMD cannot be higher than TMD), or for physiological reasons (e.g. correlation between BMD and BV/TV), we reduced the parameters for simplicity to two (linearly independent) principal components (PC1 and PC2) that explained $99\%$ of the variation on real patches.
We considered 700 real and generated patches for all resulting metrics.
Quantitative analyses of the networks were conducted by computing means and standard deviations of all considered micro-structural parameters, computed with a fixed threshold of $225~\text{mg/cm}^3$. 
We employed for all methods Tukey's range tests with $\alpha = 0.05$ between real and generated patches.

\section{Results}
\begin{table}[t!]
	\caption{Means, $\pm$ standard deviations of micro-structural parameters and $p$-values of Tukey's range test's ($\alpha = 0.05$). Bold values indicate no statistical difference.}
	\label{tab:parameters}
	\centering
	\resizebox{\linewidth}{!}
	{
		\begin{tabular}{lp{0.2cm}rlp{0.2cm}rlcp{0.2cm}rlcp{0.2cm}rlc}
			\hline
			Parameter&&\multicolumn{2}{c}{Real}&&\multicolumn{3}{c}{GAN}&&\multicolumn{3}{c}{WGAN-GP}&&\multicolumn{3}{c}{PWGAN-GP}\\
			\hline
			BMD[$\text{mg/cm}^3$]&&$123.58$ & $\pm36.21$ &&  $122.85$&$\pm31.01$ &\bm{$(99.54\%)$}&&  $120.07$&$\pm 33.04$ &\bm{$(19.01\%)$}&&  $124.76$& $\pm34.45$& \bm{$(95.04\%)$}\\
			BMD.SD[$\text{mg/cm}^3$]&&$125.97$ &$\pm22.96$ &&  $118.85$&$\pm 17.62$ &$(<0.01\%)$&&  $131.38$&$\pm 21.70$ &$(<0.01\%)$&& $123.63$& $\pm21.26$ & \bm{$(21.28\%)$}\\
			TMD[$\text{mg/cm}^3$]&&$341.94$ &$\pm30.90$ &&  $334.18$&$\pm 22.34$ &$(<0.01\%)$&&  $348.52$&$\pm 26.04$ &$(<0.01\%)$&& $338.40$& $\pm26.43$ & \bm{$(7.48\%)$}\\
			BV/TV[$\%$]&& $18.90$&$\pm7.11$ &&  $18.04$&$\pm5.45$ &\bm{$(5.82\%)$}&&  $18.67$&$\pm 6.54$ &\bm{$(95.12\%)$}&&  $18.77$&$\pm7.42$ &\bm{$(99.44\%)$}\\
			MIL[mm]&& $1.18$ &$\pm0.42$ &&  $1.13$&$\pm0.39$ &\bm{$(16.33\%)$}&&  $1.17$&$\pm 0.37$ &\bm{$(99.06\%)$}&&  $1.19$& $\pm0.40$ & \bm{$(68.20\%)$}\\
			Tb.Sp[mm]&& $0.98$ &$\pm0.43$ &&  $0.95$&$\pm0.39$ &\bm{$(37.81\%)$}&&  $0.97$&$\pm 0.37$ &\bm{$(98.45\%)$}&&  $0.99$& $\pm0.41$ & \bm{$(76.95\%)$}\\
			Tb.Th[$\mu$m]&&$197.87$&$\pm32.62$ &&  $189.65$&$\pm44.79$ &$(0.03\%)$&&  $199.21$&$\pm 47.77$ &\bm{$(96.17\%)$}&&  $199.52$& $\pm36.78$& \bm{$(70.02\%)$}\\
			\hline
	\end{tabular}}
\end{table}
\begin{figure*}[t!]
	\centering
	\includegraphics[width=\linewidth]{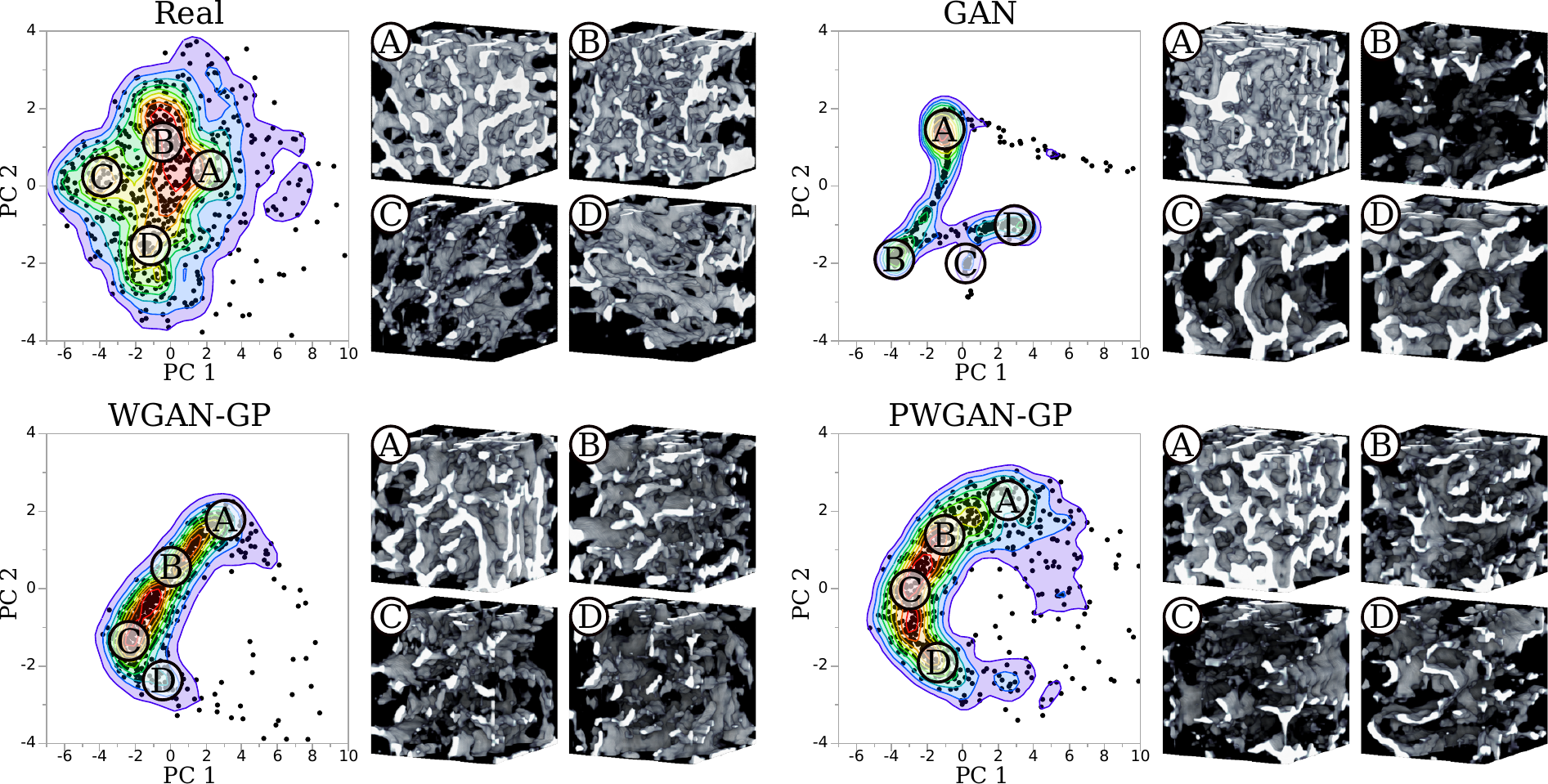}
	\caption{Distributions of principal components of 700 randomly selected patches, and each four 3D renderings of representative samples of real and generated sets of three different architectures.} 
	\label{fig:pc1pc2}
\end{figure*}
Figure~\ref{fig:pc1pc2} shows the distribution of real and generated patches.
The upper left plot indicates the target distribution, noticeably GAN differs most and PWGAN-GP least from the real distribution.
Besides that all neural networks still underpopulated certain areas in parameter space, e.g. PWGAN-GP at (PC1,PC2) = (2,-1), single generated patches were visually difficult to distinguish from real patches.
Table~\ref{tab:parameters} shows statistics of micro-structural parameters. 
Only PWGAN-GP (right column) was indistinguishable on all considered statistics, while WGAN-GP was statistically different from real data on parameters BMD.SD and TMD, and GAN additionally on Tb.Th.
\begin{figure*}[t!]
	\centering
	\includegraphics[width=\linewidth]{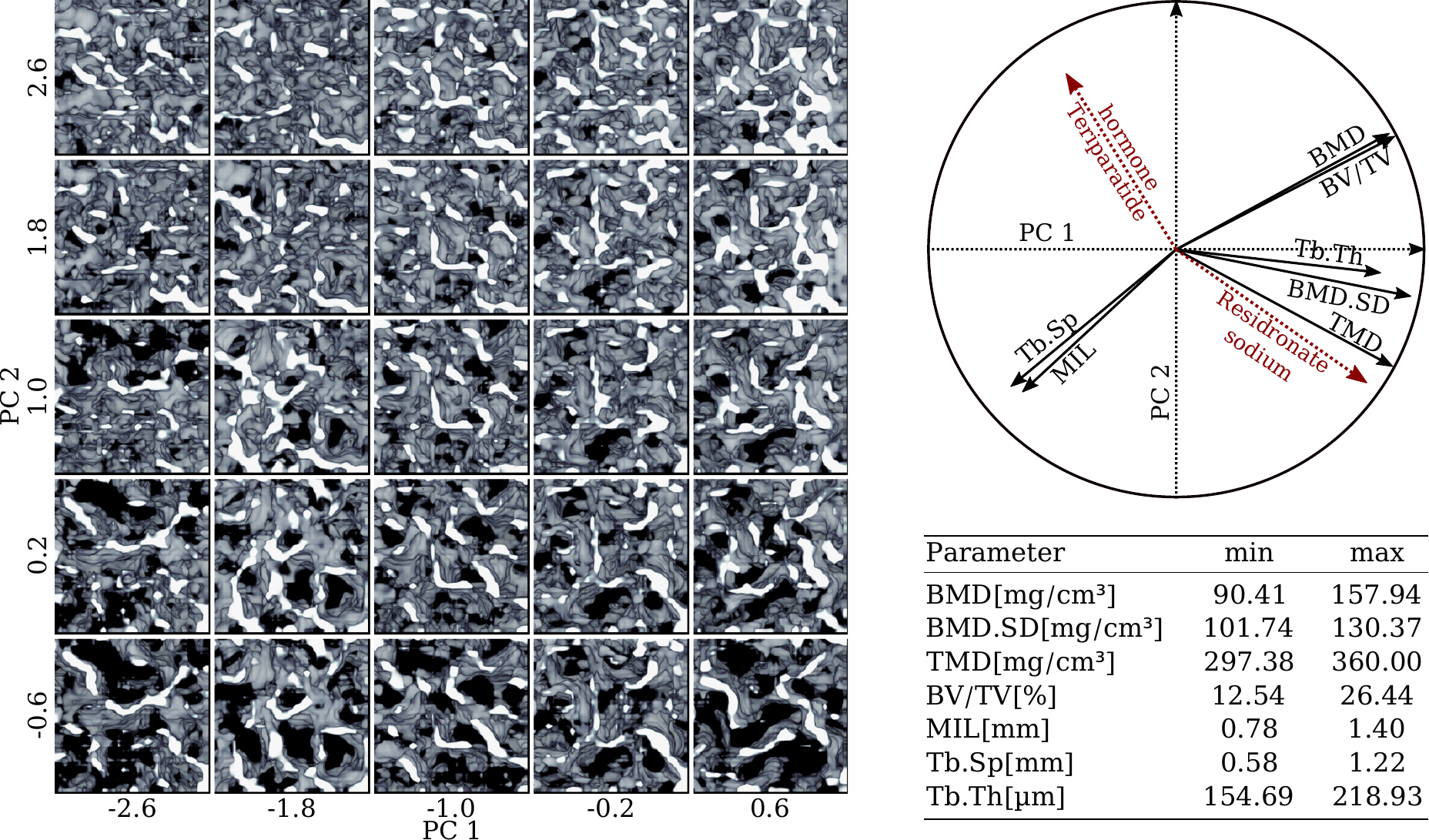}
	\caption{Generated samples with varying micro-structural parameters but fixed content. Right: Direction of change of structural parameters and two specific drugs to strengthen bone (Fig.~\ref{fig:treatment}), range of measured parameters.} 
	\label{fig:5x5}
\end{figure*}
Figure~\ref{fig:5x5} shows the application of the neural style transfer method on 25 bone samples with different micro-structural parameters. 
The patches are variations of a single patch which was directly generated by the GAN with principal components close to the one in the center, thereby keeping bone micro structure as fixed as possible by simultaneously fitting the individual target structural parameters.
The circle on the top right shows the change of principal components for each micro-structural parameter.
Also treatment effects of drugs for osteoporosis therapy can be expressed in such a way, as shown for Residronate and Teriparatide.
Therefore, we used reported treatment effects of the examined structural parameters~\cite{Gluer2013Comparative}.

\section{Discussion and Conclusion}
We implemented a method to generate realistic \emph{in-silico} patches of bone micro-structure with defined micro-structural parameters.  
Qualitative analyses showed high similarity with real bone.
Micro-structural parameters were indistinguishable between generated and real patches, which is particularly important since the loss-function of the neural network was not explicitly considering these statistics, hence further micro-structural characteristics might be modeled correctly as well.

We still see potential to improve our method, i.e. to closer align the distributions of generated and real patches (Fig.~\ref{fig:pc1pc2}). 
Latent vectors that are an explicit combination of style-generating and structural variables might be worth of consideration.
Furthermore, network architectures applied for similar problems like the generation of blood vessels~\cite{Russ2019synthesis}, lung nodules~\cite{onishi2019automated}, liver lesions~\cite{frid2018gan} or other applications~\cite{sorin2020creating} seem to be promising alternatives to consider.
Our method generated only patches of 5mm (or $32^3$ voxels).
We tried also to train for patches with 10mm (or $64^3$) thereby using a larger number of epochs and latent dimensions, but application on larger scale became unstable according to the distribution criterion.
An alternative might be a generative model to produce continuous bone structures which required however a more complex logic to maintain a smooth continuum of bone structural parameters.

\begin{figure*}[t!]
	\centering
	\includegraphics[width=\linewidth]{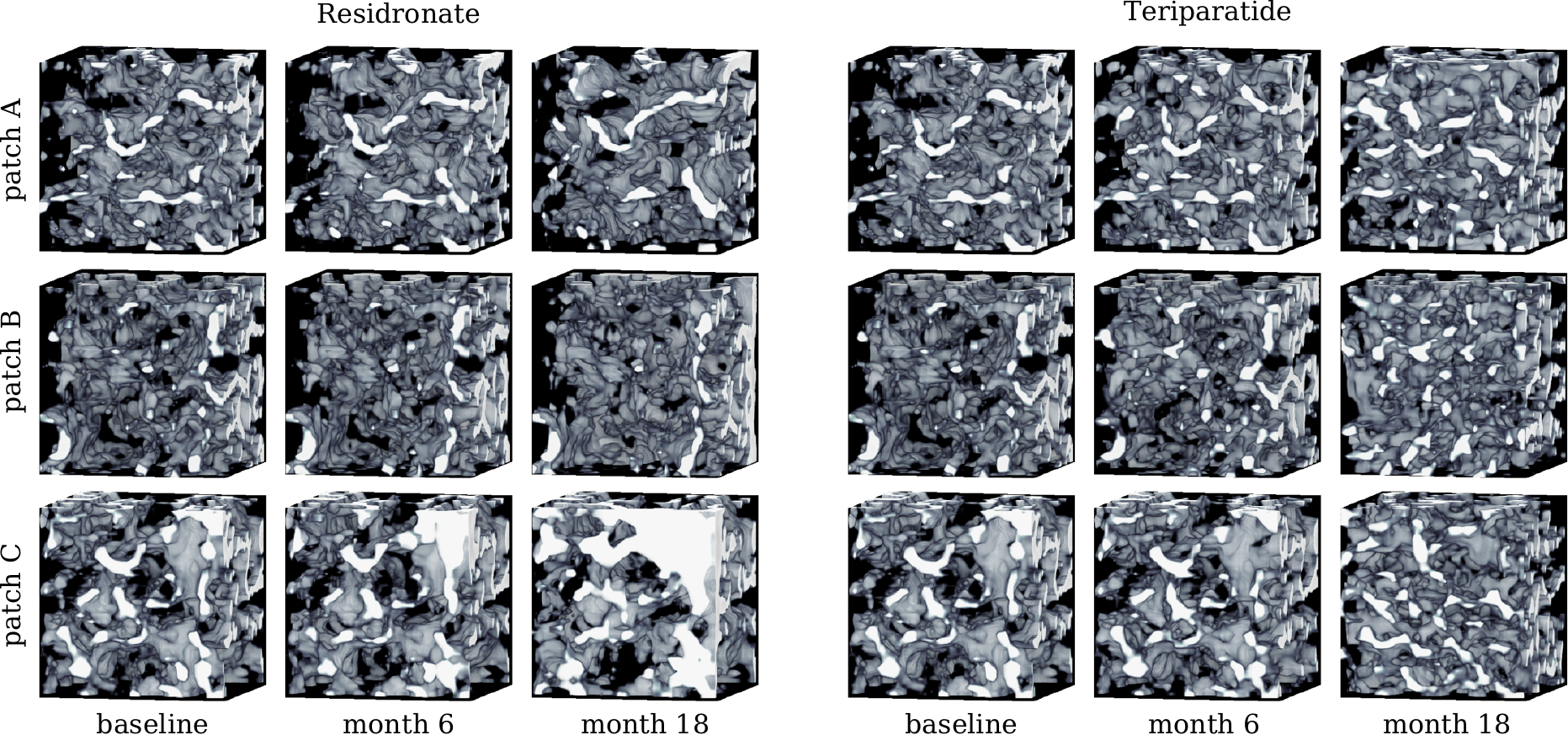}
	\caption{Simulated bone structure under Residronate and Teriparatide treatment after 6 and 18 months. While Residronate (left) is known to calcify existing bone only, Teripartide (right) is able to form new bone.} 
	\label{fig:treatment}
\end{figure*}
We foresee our method being useful in a number of applications.
For instance, the generation of bone structures for developing and testing of new micro-structural parameters that enhance our current understanding of bone stability.
Also, the simulation of micro-structural changes by bone-forming drugs or immobility- or age related bone-reduction is still an open issue. 
Since micro-structural properties of specific treatments for osteoporosis are generally known, the content-preserving method to vary micro-structural parameters can be used to simulate bone micro-structure after a specific treatment.
Such simulations are shown in Fig.~\ref{fig:treatment} based on reported treatment effects of the bisphosphonate Residronate and the parathyroid hormone Teriparatide using reported effects on BMD, BV/TV and TMD~\cite{Gluer2013Comparative}.
The proposed method could potentially be refined by incorporating differentiable formulas of additional structural parameters, such as Tb.Sp, Tb.Th, the anisotropy and elongation indices.
The simulated micro-structure might also serve to compute minimum bone failure load with the finite-element-method or the mean-intercept-length tensor~\cite{MorenoFabric2014}, which both require exact models of the bone.

\noindent\\
\textbf{Acknowledgements}\\
We thank C.-C.~Gl\"uer for providing the phantoms. 
This study was supported by Agencia Nacional de Promoci\'on Cient\'ifica y Tecnol\'ogica, Argentina (PICT 2017-1731), PID UTN 2018 (SIUTNBA0005139), PID UTN 2019 (SIUTNBA0005534), and NVIDIA GPU hardware grant that supported this research with the donation of two Titan Xp graphic cards.

\bibliographystyle{splncs04}
\bibliography{Bib1691}

\begin{thebibliography}{10}
\providecommand{\url}[1]{\texttt{#1}}
\providecommand{\urlprefix}{URL }
\providecommand{\doi}[1]{https://doi.org/#1}

\bibitem{arjovsky2017wasserstein}
Arjovsky, M., Chintala, S., Bottou, L.: Wasserstein {GAN}. arXiv preprint
  arXiv:1701.07875  (2017)

\bibitem{frid2018gan}
Frid-Adar, M., Diamant, I., Klang, E., Amitai, M., Goldberger, J., Greenspan,
  H.: {GAN}-based synthetic medical image augmentation for increased {CNN}
  performance in liver lesion classification. Neurocomputing  \textbf{321},
  321--331 (2018)

\bibitem{gatys2015neural}
Gatys, L.A., Ecker, A.S., Bethge, M.: A neural algorithm of artistic style.
  arXiv preprint arXiv:1508.06576  (2015)

\bibitem{Gluer2013Comparative}
Gl{\"u}er, C.C., Marin, F., Ringe, J.D., Hawkins, F., M{\"o}ricke, R.,
  Papaioannu, N., Farahmand, P., Minisola, S., Mart\'{\i}nez, G., Nolla, J.M.:
  Comparative effects of {Teriparatide and Risedronate} in
  glucocorticoid-induced osteoporosis in men: 18-month results of the
  {EuroGIOPs} trial. J. Bone Miner. Res.  \textbf{28},  1355--1368 (2013)

\bibitem{goodfellow2016deep}
Goodfellow, I., Bengio, Y., Courville, A.: Deep learning. MIT press (2016)

\bibitem{gulrajani2017improved}
Gulrajani, I., Ahmed, F., Arjovsky, M., Dumoulin, V., Courville, A.C.: Improved
  training of {Wasserstein GANs}. In: Advances In Neural Information Processing
  Systems. pp. 5767--5777 (2017)

\bibitem{karam2019fast}
Karam, C., Sugimoto, K., Hirakawa, K.: Fast convolutional distance transform.
  IEEE Signal Process. Lett.  \textbf{26}(6),  853--857 (2019)

\bibitem{karras2017progressive}
Karras, T., Aila, T., Laine, S., Lehtinen, J.: Progressive growing of {GANs}
  for improved quality, stability, and variation. arXiv preprint
  arXiv:1710.10196  (2017)

\bibitem{kingma2014adam}
Kingma, D.P., Ba, J.: Adam: A method for stochastic optimization. In: Int.
  Conf. on Learning Representations. pp. 1--15 (2015)

\bibitem{mirza2014conditional}
Mirza, M., Osindero, S.: Conditional generative adversarial nets. arXiv
  preprint arXiv:1411.1784  (2014)

\bibitem{Moreno2012evaluation}
Moreno, R., Borga, M., Smedby, {\"O}.: Evaluation of the plate-rod model
  assumption of trabecular bone. In: IEEE Int. Symposium on Biomedical Imaging.
  pp. 470--473 (2012)

\bibitem{MorenoFabric2014}
Moreno, R., Borga, M., Smedby, {\"O}.: Techniques for computing fabric tensors:
  A review. In: Westin, C.F., Vilanova, A., Burgeth, B. (eds.) Visualization
  and Processing of Tensors and Higher Order Descriptors for Multi-Valued Data.
  pp. 271--292. Springer (2014)

\bibitem{onishi2019automated}
Onishi, Y., Teramoto, A., Tsujimoto, M., Tsukamoto, T., Saito, K., Toyama, H.,
  Imaizumi, K., Fujita, H.: Automated pulmonary nodule classification in
  computed tomography images using a deep convolutional neural network trained
  by generative adversarial networks. {BioMed} Research Int.  \textbf{2019}
  (2019)

\bibitem{pena2019development}
Pe{\~n}a-Sol{\'o}rzano, C.A., Albrecht, D.W., Paganin, D.M., Harris, P.C.,
  Hall, C.J., Bassed, R.B., Dimmock, M.R.: Development of a simple numerical
  model for trabecular bone structures. Med. Phys.  \textbf{46}(4),  1766--1776
  (2019)

\bibitem{Russ2019synthesis}
Russ, T., Goerttler, S., Schnurr, A.K., Bauer, D.F., Hatamikia, S., Schad,
  L.R., Z{\"o}llner, F.G., Chung, K.: Synthesis of {CT} images from digital
  body phantoms using {CycleGAN}. Int. J. Comput. Assist. Radiol. Surg.
  \textbf{14}(10),  1741--1750 (2019)

\bibitem{sorin2020creating}
Sorin, V., Barash, Y., Konen, E., Klang, E.: Creating artificial images for
  radiology applications using generative adversarial networks {(GANs)}--a
  systematic review. Acad. Radiol.  (2020)

\bibitem{Thomsen2017PHD}
Thomsen, F.: Medical {3D} image processing applied to computed tomography and
  magnetic resonance imaging. Ph.D. thesis, Universidad Nacional del Sur,
  Bah{\'i}a Blanca, Argentina (2017)

\bibitem{Thomsen2016SI}
Thomsen, F., Pe{\~n}a, J., Delrieux, C., Gl{\"u}er, C.C.: {Structural Insight
  v3}: A stand-alone program for micro structural analysis of computed
  tomography volumes. In: Congreso Argentino de Inform\'atica y Salud (2016)

\bibitem{Thomsen2016LFD}
Thomsen, F., Pe{\~n}a, J., Lu, Y., Huber, G., Morlock, M., Gl{\"u}er, C.C.,
  Delrieux, C.: A new algorithm for estimating the rod volume fraction and the
  trabecular thickness from in vivo computed tomography. Med. Phys.
  \textbf{43}(12),  6598--6607 (2016)

\bibitem{yazici2018unusual}
Yaz{\i}c{\i}, Y., Foo, C.S., Winkler, S., Yap, K.H., Piliouras, G.,
  Chandrasekhar, V.: The unusual effectiveness of averaging in {GAN} training.
  arXiv preprint arXiv:1806.04498  (2018)

\end{thebibliography}

\section*{Supplemental Document - Computation of Differentiable Structural Parameters}
\subsection*{Standard formulas}
We briefly describe the differentiable formulas of the structural parameters used in the style transfer optimization.
We implemented structural parameters that resemble common parameters bone mineral density (BMD), its standard deviation (BMD.SD), bone volume ratio (BVTV) and tissue mineral density (TMD).
BMD is the average density of the calibrated input volume $\boldsymbol{x}$.
$\text{BVTV}_t$ is the ratio of bone to total volume when segmenting with threshold~$t$ and $\text{TMD}_t$ is the density of the segmented bone. 
The standard formulas read
\begin{eqnarray}
&\text{BMD}(\boldsymbol{x})&= \; \frac{1}{n} \sum \boldsymbol{x}_i,\\
&\text{BMD.SD}(\boldsymbol{x}) &= \sqrt{\frac{\sum\boldsymbol{x}_i^2 - \frac{1}{n}(\sum\boldsymbol{x}_i)^2}{n-1}},\\
&\text{BVTV}_t(\boldsymbol{x})&= \; \frac{1}{n} \sum \HS(\boldsymbol{x}_i-t) \text{and}\\
&\text{TMD}_t(\boldsymbol{x})&= \; \frac{\sum \HS(\boldsymbol{x}_i-t) \; \boldsymbol{x}_i}{\sum \HS(\boldsymbol{x}_i-t)},
\end{eqnarray}
with $\HS(\cdot)$ the Heaviside function, $i$ a voxel index and $n$ the total number of voxels.

\subsection*{Differentiable formulas}
For the use in the neural style transfer optimizer we derived new formulas that 1)~allow computation on a very restricted local VoI of $32^3$ voxels instead of a global VoI per-vertebra, and 2)~are differentiable to become part of the operator $P(\boldsymbol{x})$. 
Thus we reimplemented the threshold dependent parameters TMD and BV/TV. 
Smoothness was achieved by defining parameters that are strictly monotone in~$t$. 
We defined a)~the softplus function $\softplus{\cdot}$ that is a smooth version of the rectified linear unit function $\max\{0,\cdot\}$ parameterized with a fuzziness factor $\epsilon$ and with $\softplus{a}\approx a$ for $a>3.5\,\epsilon$,
and b)~a fuzzy binarization $\fHS(a-t)$, a sigmoid version of the Heaviside function $\HS(a-t)$ with $\sigma>0$:
\begin{eqnarray}
&\softplus{a} &= \epsilon \ln\left(1+\exp\left(\frac{a}{\epsilon}-1\right)\right)+\epsilon,\\
&\fHS(a-t) &=\left(1+\exp\left(\frac{t-a}{\sigma}\right)\right)^{-1}.
\label{Eq:SoftMax}
\end{eqnarray}
Smooth structural parameters read then
\begin{eqnarray}
&\text{BVTV}^{\ast}_{t}(\boldsymbol{x})&=\frac{1}{n} \sum \fHS(\boldsymbol{x}_i-t),\\
&\text{TMD}^{\ast}_{t}(\boldsymbol{x})&=\frac{\sum \fHS(\boldsymbol{x}_i-t) \; \boldsymbol{x}_i}{\softplus{\sum \fHS(\boldsymbol{x}_i-t)}}
\end{eqnarray} 
with $\epsilon=10^{-4}$ a small number and $\sigma=10~\text{mg/cm}^3$ a fuzzy scale.
Finally the term $P(\cdot)$ as used in the paper reads
\begin{equation}
P(\boldsymbol{x}) = \langle \alpha_1\,\text{BMD}(\boldsymbol{x}),\; \alpha_2\,\text{BMD.SD}(\boldsymbol{x}),\; \alpha_3\,\text{BVTV}^{\ast}_{t}(\boldsymbol{x}),\; \alpha_4\,\text{TMD}^{\ast}_{t}(\boldsymbol{x}) \rangle 
\end{equation}
with $t=225~\text{mg/cm}^3$ and $\alpha_1,\ldots,\alpha_4$ normalization or weighting factors for each parameter.  

\end{document}